\documentclass[12pt,preprint]{aastex}

\usepackage{longtable}
\usepackage{epstopdf}  
\usepackage{ifthen}
\usepackage{rotating}
\usepackage{subfigure} 

\received{February 7, 2011}
\accepted{March 28, 2011}
\shorttitle{CN Morphology Studies of 103P/Hartley 2}
\shortauthors{Knight et al.}
\journalid{The Astronomical Journal}{}
\articleid{}{}

\begin{document}

\title{CN Morphology Studies of Comet 103P/Hartley 2}

\author{Matthew M. Knight \altaffilmark{1,2} and David G. Schleicher \altaffilmark{2}}

\altaffiltext{1}{Contacting author: knight@lowell.edu.}
\altaffiltext{2}{Lowell Observatory, 1400 W. Mars Hill Rd, Flagstaff, AZ 86001, U.S.A.}

\begin{abstract}
We report on narrowband CN imaging of Comet 103P/Hartley 2 obtained at Lowell Observatory on 39 nights from 2010 July until 2011 January. We observed two features, one generally to the north and the other generally to the south. The CN morphology varied during the apparition: no morphology was seen in July; in August and September the northern feature dominated and appeared as a mostly face-on spiral; in October, November, and December the northern and southern features were roughly equal in brightness and looked like more side-on corkscrews; in January the southern feature was dominant but the morphology was indistinct due to very low signal. The morphology changed smoothly during each night and similar morphology was seen from night to night. However, the morphology did not exactly repeat each rotation cycle, suggesting that there is a small non-principal axis rotation.
Based on the repetition of the morphology, we find evidence that the fundamental rotation period was increasing: 16.7 hr from August 13--17, 17.2 hr from September 10--13, 18.2 hr from October 12--19, and 18.7 hr from October 31--November 7. We conducted Monte Carlo jet modeling to constrain the pole orientation and locations of the active regions based on the observed morphology. Our preliminary, self-consistent pole solution has an obliquity of 10$^\circ$ relative to the comet's orbital plane (i.e., it is centered near RA = 257$^\circ$ and Dec=$+$67$^\circ$ with an uncertainty around this position of about 15$^\circ$) and has two mid-latitude sources, one in each hemisphere.
\end{abstract}


\section{Introduction}
\label{sec:intro}
As the subject of a flyby by NASA's EPOXI spacecraft on 2010 November 4 \citep{ahearn11}, the 2010 apparition of Comet 103P/Hartley 2 provided a rare opportunity to link the macroscopic information gleaned from the Earth with in situ observations from the spacecraft. Very favorable viewing geometry from the Earth allowed observations of Hartley 2 over a long time baseline with high spatial resolution. As a result, Hartley 2 was widely studied and the ensemble of observations will yield unprecedented insight into the composition and rotational behavior of Hartley 2 (see the overview by \citet{meech11}). 

Hartley 2 is a Jupiter family comet discovered by Malcolm Hartley in 1986 \citep{iauc4197}. While its perihelion distance is currently 1.06 AU, it had an Earth-crossing orbit in the recent past. It reached perihelion on the far side of the Sun in 1985, but the 1991 and 1997 apparitions were reasonable and studies were conducted by a number of researchers during these apparitions. In 2004, it again reached perihelion on the far side of the Sun and was not studied extensively. The 2010 apparition was the best since its discovery, with Hartley 2 passing within 0.12 AU of the Earth on October 20, only a week before perihelion (October 28), and two weeks before the EPOXI encounter. 

Previous studies have generally concentrated on either composition, level of activity, or nucleus size. Compositional studies by \citet{weaver94}, \citet{ahearn95}, and \citet{colangeli99} generally found a normal composition. Numerous studies found it to be highly active and could only set upper limits on the nucleus size \citep{licandro00,lowry01,snodgrass06,tancredi06,mazzotta08,ferrin10b}, while studies by \citet{groussin04} and \citet{lisse09} found a sub-kilometer nucleus. The forthcoming results from the EPOXI flyby \citep{ahearn11} should definitively measure the size of the nucleus and may place constraints on the composition and level of activity. 

In preparation for the EPOXI flyby, \citet{meech09} found a nucleus rotation period near 16.6 hr using a number of 8--10 m telescopes and {\it HST} in March through July of 2009. Widespread observations began in earnest in mid-2010. We discovered CN morphology whose variation repeated with a period of 16.6 $\pm$ 0.5 hr from August 13 to 17 \citep{iauc9163a}. The presence of CN morphology was confirmed by \citet{iauc9178} who found a periodicity near 17.1 hr from September 1 to 3 and near 17.6 hr from September 30 to October 4. \citet{iauc9179a} obtained radar observations at Arecibo Observatory from October 24 to 27, reporting a rotation period of 18.1 $\pm$ 0.3 hr and a ``highly elongated, bilobate object with a long-axis dimension of at least 2.2 km.'' \citet{cbet2589} monitored Hartley 2 from October 29 until December 7, finding evidence for a lengthening rotation period consistent with earlier reported values. Slight differences in the CN morphology from cycle to cycle were reported by \citet{iauc9178} and \citet{iauc9175}, suggesting the comet may have a small complex component of the rotation. \citet{cbet2515c} reported very strong variability of the J=4--3 transition on the timescale of hours from September 29 to October 1 and October 16 to 19, suggesting ``a much longer rotation period [than 17.6 hr], significant excitation of the rotation state, or a non-periodic nature of the detected variability.'' Finally, \citet{iauc9177a} reported unusual CN behavior from September 9 to 17 as the CN emission measured from EPOXI increased then decreased relative to the dust continuum.


We observed Hartley 2 monthly from 2010 July until 2011 January using both broadband and narrowband filters \citep{farnham00}. This paper is restricted to discussion of the CN images, which comprised the plurality of our images. Relative to the other gas species observable in visible bandpasses (most frequently OH, C$_3$, and C$_2$), CN typically has the best combination of brightness and contrast to the underlying dust continuum. The higher velocity of gas outflow ensures that older CN gas does not remain in the field of view for long, unlike dust. Thus, it is easier to observe morphology that traces the nucleus rotation in CN images than in the brighter dust images (broadband R and narrowband continuum). A second paper is planned to discuss the imaging obtained in filters other than CN as well as complementary narrowband photometry. 

CN coma morphology has become a valuable tool for inferring information about cometary nuclei. Since CN morphology was first observed in 1P/Halley by \citet{ahearn86}, it has been used to study other comets such as C/2004 Q2 Machholz \citep{farnham07a} and 8P/Tuttle \citep{waniak09}. The source of CN in cometary comae is uncertain. CN is a daughter species; it is not released directly from the nucleus but is produced from either the photodissociation of a parent molecule or is emitted directly from grains in the coma (e.g. \citet{bockeleemorvan85,fray05}). While HCN is probably the largest single source of CN, the production rates of HCN relative to CN vary from comet to comet, with some comets 
having production rates consistent with all the CN coming from HCN while other comets 
require an additional source of CN (\citet{fray05} and references therein). 
The published HCN production rates for Hartley 2 \citep{iauc9171,iauc9180c} are about $\sim$10\% lower than our unpublished CN production rates, suggesting that HCN is the dominant source of CN in Hartley 2. 

The layout of this paper is as follows. In Section~\ref{sec:reductions} we summarize the observations and data reductions. In Section~\ref{sec:results} we discuss the CN morphology seen at each observing epoch, discuss the evolution of the morphology, consider evidence for non-principal axis rotation of the nucleus, determine the rotation period of the nucleus throughout our observations, attempt to replicate the data with Monte Carlo jet modeling, and characterize the comet at the time of the EPOXI flyby. We summarize our results in Section~\ref{sec:summary}.

\section{Observations and Reductions}
\label{sec:reductions}
We observed Hartley 2 for a total of 39 nights using the Hall 1.1-m and the 0.8-m telescopes at Lowell Observatory (Table~\ref{t:obs_summary}). To increase the signal to noise and lessen the effects of long star trails we typically observed in sets of 3--10 images in the same filter. The comet was monitored throughout the night for most 1.1-m nights while all 0.8-m nights consisted of one to five sets of snapshot observations and were acquired robotically. The 1.1-m images were obtained using an e2v CCD231-84 chip with 4K$\times$4K pixels which, when binned 2$\times$2 at the telescope, produced a pixel scale of 0.74 arcsec. We rebinned these images by an additional factor of two during processing to increase the effective signal to noise, resulting in a final pixel scale of 1.48 arcsec. The 0.8-m images were obtained using an e2v CCD42-40 chip with 2K$\times$2K pixels and a pixel scale of 0.46 arcsec. The images were also binned by a factor of two during processing to increase the signal to noise, resulting in a final pixel scale of 0.92 arcsec. We primarily imaged the comet with broadband R and narrowband CN filters, but additional HB narrowband filters \citep{farnham00} were occasionally used as well. All 1.1-m images were guided at the comet's rate of motion while all 0.8-m images were tracked at the comet's rate of motion. 


The data were reduced using standard bias and flat field techniques. 
As the current analysis considers only the CN morphology, we did not apply absolute calibrations (and on non-photometric nights we could not apply absolute calibrations). To verify that this was appropriate, we fully processed images on August 13 and November 3.
These cases represent the extremes of our observations: low activity, low signal to noise, and low spatial resolution in August and high activity, high signal to noise, and high spatial resolution in November. CN images have contamination from C$_3$ gas and dust continuum. In order to get a pure CN image we used fully processed C$_3$ and blue continuum images and removed their relative contributions in the CN bandpass using the methodology of \citet{farnham00}. The resulting pure CN images exhibited the same morphology as the contaminated CN images we used in our analysis. The maximum contamination occurs at the nucleus and drops with projected distance because the C$_3$ has a shorter lifetime than CN and the dust falls off more rapidly ($\rho$$^{-1}$ or steeper) than CN. The C$_3$ morphology was similar to the CN morphology, although it appeared more diffuse. C$_3$ contributed less than 20\% of the signal beyond $\sim$4 pixels from the nucleus ($\sim$500 km) in November and less than 10\% of the signal beyond $\sim$5 pixels from the nucleus ($\sim$3000 km) in August. Thus, C$_3$ contamination did not affect the CN morphology. While the dust morphology differed from the CN morphology (the dominant dust feature was the tail which pointed $\sim$180$^\circ$ from the PA of the Sun given in Column 8 of Table~\ref{t:obs_summary}), its contribution fell off rapidly. The dust contributed 
less than 5\% of the signal beyond $\sim$4 pixels from the nucleus in both November ($\sim$650 km) and August ($\sim$2000 km). 
The dust tail was not evident in our CN images, confirming that over the region of study there was not significant contamination from the dust continuum.

In order to better see the features that are hidden by the bulk coma brightness, we enhanced the bias-subtracted and flat-fielded CN images using several standard techniques. Our primary method was division by an azimuthal median profile, although other enhancements were also tested: subtraction of an azimuthal median profile, division/subtraction of a $\rho$$^{-1}$ profile, division/subtraction of a rotationally phased image, azimuthal renormalization, Laplace filter, and unsharp mask (see e.g. \citet{schleicher04}). The azimuthal median division was utilized most frequently and is shown here because it enhances more subtle features of the coma without introducing artifacts that are not present in the original image, and does not depend on a kernel size. 
This requires accurate centroiding on the nucleus, and images that are out of focus or have very strong jets can pull the centroid away from the nucleus. We found that for positions within 2 binned pixels of the nominal centroid the enhanced images looked similar, but larger deviations produced increasingly different morphology. We are confident that all images used for analysis were centroided to an accuracy of better than 1 pixel. While each enhancement method emphasizes different characteristics, the features discussed in the subsequent section were present after several different enhancement techniques were applied, engendering confidence that the features we describe are, in fact, real.

\section{Results}
\label{sec:results}
\subsection{CN Morphology Overview}
\label{sec:morphology}
Representative enhanced CN images are shown in Figure~\ref{fig:cn_images} for each night of imaging data except 2010 July 19 when the comet was too faint to detect morphology within the CN emission. All images are oriented so north is up and east is to the left, and each image is $\sim$50,000 km across at the comet. Three images per night are typically shown and images are usually spaced 3--5 hr apart to show the morphology throughout a night and from night to night. However, nights with less temporal coverage (due to snapshot observations, poor weather, telescope problems, or simply because the comet was only available for less than 5 hr) have fewer than three images displayed. In these cases, blank images are inserted to indicate at which point in a night data were not obtained. Note that the morphology changed smoothly throughout each night; thus even on under-sampled nights we are confident that we know the general morphology between our images. Stars appear as trails in most images, however, a magnitude 3.6 binary star (HD 217675 or Omicron Andromedae) passed within $\sim$10 arcmin of the comet on 2010 September 9 and swamped the coma signal. Thus the bright white feature seen on this night towards the north, then northwest, then southwest is not due to coma morphology but to background star contamination.


The enhancements reveal that the coma is not uniform and suggest that the nucleus is not 100\% active as had been previously suggested \citep{groussin04}. In August and September, the morphology appears as a nearly face-on spiral rotating clockwise. The spiral appears tilted to our line of sight so the majority of the signal is to the northwest. During October, November, and December features are seen in both the north and south with the northern feature generally brighter than the southern feature. In January only a southern feature was evident, although we note that the temporal coverage was limited and a northern feature may have been evident during unobserved phases. The morphology was generally straighter from October onward than in August and September, and rotational motion was less obvious.

\subsection{Evolution of Morphology}
\label{sec:morph_evolution}
As is evident in Figure~\ref{fig:cn_images}, the CN morphology evolved during the apparition. The northern feature appeared as a nearly face-on spiral in August and September suggesting that the Earth was inside the cone of material swept out as the nucleus rotated. The spiral appears to be inclined to our line of sight such that most of the signal is in the north and the feature sweeps through our line of sight slightly to the south. While the viewing geometry was relatively similar in August and September, it changed dramatically in October and November as the comet reached perigee, passing only 0.12 AU from the Earth. During our October observations, the northern feature appeared to be at an angle to the plane of the sky, at times making part of a face-on spiral (see image 101016 8:19 in Figure~\ref{fig:cn_images}) but at other times appearing to be a side-on corkscrew (101017 9:05). By early November, the feature appeared to be nearly side-on all of the time with some evidence of a corkscrew shape (101103 6:53). The northern feature appeared somewhat less side-on in December, and was not evident in January (although rotational coverage was limited). By December and January, the viewing geometry was no longer changing rapidly and our view was from roughly the opposite side of the comet as compared to August and September. The large change in viewing geometry during the apparition allowed us to see the morphology from many view points and helped constrain our preliminary model (discussed in subsection~\ref{sec:modeling}).

A southern feature appeared during the apparition. This feature was not evident in August and was only marginally evident in September, but was quite evident in October, November, December, and January. The appearance of this second feature suggests that there is a seasonal change of activity on the nucleus of Hartley 2. Furthermore, the strength of the two features relative to each other appears to vary during the rotation cycle. However, as we discovered when conducting our Monte Carlo jet modeling (subsection~\ref{sec:modeling}), these phenomena can be explained by the overlap (from our observing perspective) of material from two source regions, and may not be indicative of a seasonal change.

\begin{figure}
  \centering
  \epsscale{0.35}
  \plotone{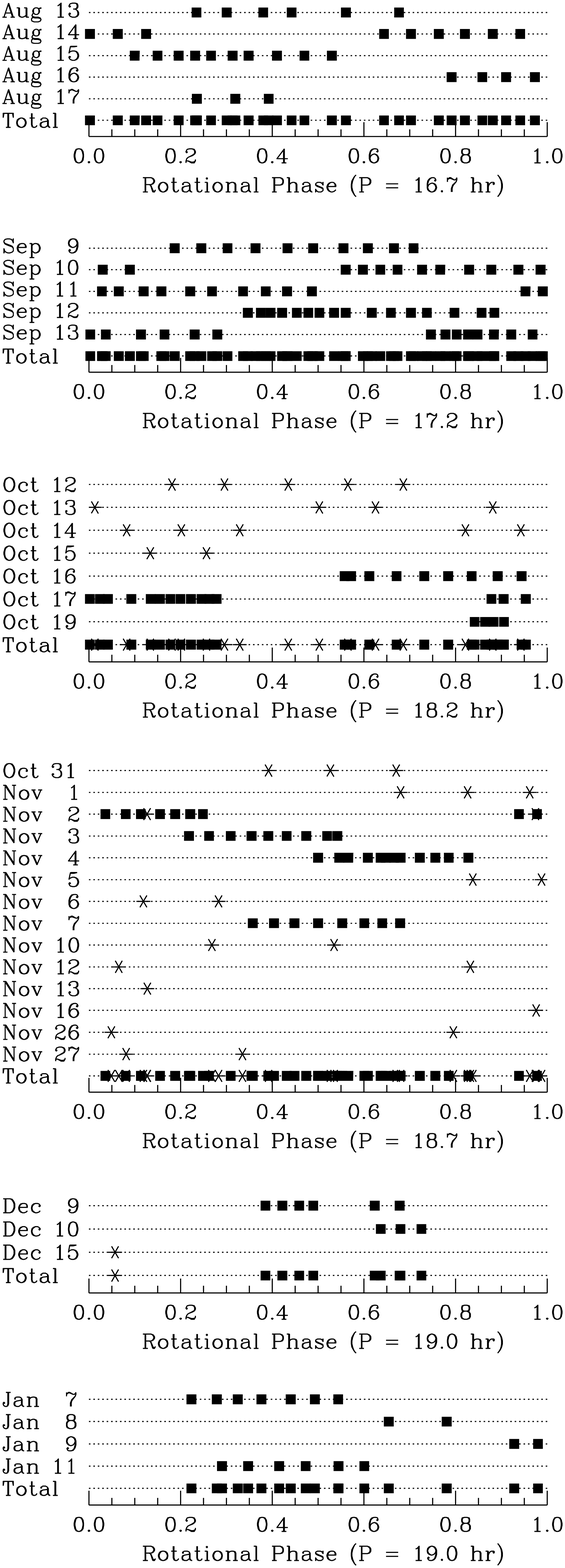}    
  \caption[Rotation phase coverage]{Rotational phase coverage during each observing run. The times of observations are phased to 16.7 hr for August 13--17, 17.2 hr for September 9--13, 18.2 hr for October 12--19, 18.7 hr for October 31--November 27, and 19.0 hr for December and January. The total phase coverage for each observing run is given at the bottom of each run's plot. The 1.1-m data are plotted as filled squares while 0.8-m data are plotted as asterisks. Because the rotation period is changing throughout the apparition, zero phase was set at 0:00 UT on the first night of each observing run.}
  \label{fig:phase_coverage}
\end{figure}

The general alignment of the morphology changed during the apparition. Measuring the alignment of the morphology can help constrain the pole orientation (subsection~\ref{sec:modeling}) since the jet of material from a source that is active throughout a rotation cycle will sweep out a cone centered on the rotation pole. When the cone is viewed from the side (i.e., the Earth is outside of the cone), the position angle (PA) of the center will be reasonably well defined, whereas if the Earth is within the cone, the morphology will appear as a face-on spiral, and defining a central PA is difficult.
We combined rotationally phased images during each of our 1.1-m observing runs to determine the PA of the source during the run. 
Images were phased to the rotation periods that will be discussed in subsection~\ref{sec:rot_period}, were evenly spaced in phase, and were rescaled to account for obscuration on non-photometric nights.
The PAs (measured counterclockwise from north) from August and September were uncertain as the Earth appeared to be inside the cone, although the morphology was predominantly in the north. We measured PAs of 350$^\circ$ $\pm$ 20$^\circ$ from August 13--17 and 335$^\circ$ $\pm$ 20$^\circ$ from September 10--13. The morphology appeared to be more side-on in October, November, and December. While this allowed us to constrain the PA better than in August and September, we had incomplete phase coverage in October and December, so these PAs have larger uncertainties than November. The northern feature had PAs of 350$^\circ$ $\pm$ 10$^\circ$ from October 16--17, 10$^\circ$ $\pm$ 5$^\circ$ from November 2--4, and 55$^\circ$ $\pm$ 15$^\circ$ from December 9--10. We measured PAs of the southern feature of 145$^\circ$ $\pm$ 10$^\circ$ (October), 190$^\circ$ $\pm$ 5$^\circ$ (November), and 215$^\circ$ $\pm$ 15$^\circ$ (December). The northern and southern PAs in October and December are not $\sim$180$^\circ$ apart which we believe to be due to day/night variation in intensity of the northern feature causing the average center of brightness to be eastward of the projected pole.
The data were insufficient to determine a PA in January. Note that the PAs given above differ somewhat from those given in \citet{iauc9175} because of a centroiding problem in our preliminary analysis. Of our five PA measurements, we consider the November measurement to be the strongest constraint on the pole orientation because it had the best rotational coverage (see Figure~\ref{fig:phase_coverage}) and the two features appeared to be viewed most side-on. 



\subsection{Rotation Period}
\label{sec:rot_period}
\begin{figure}
  \centering
  \epsscale{0.7}
  \plotone{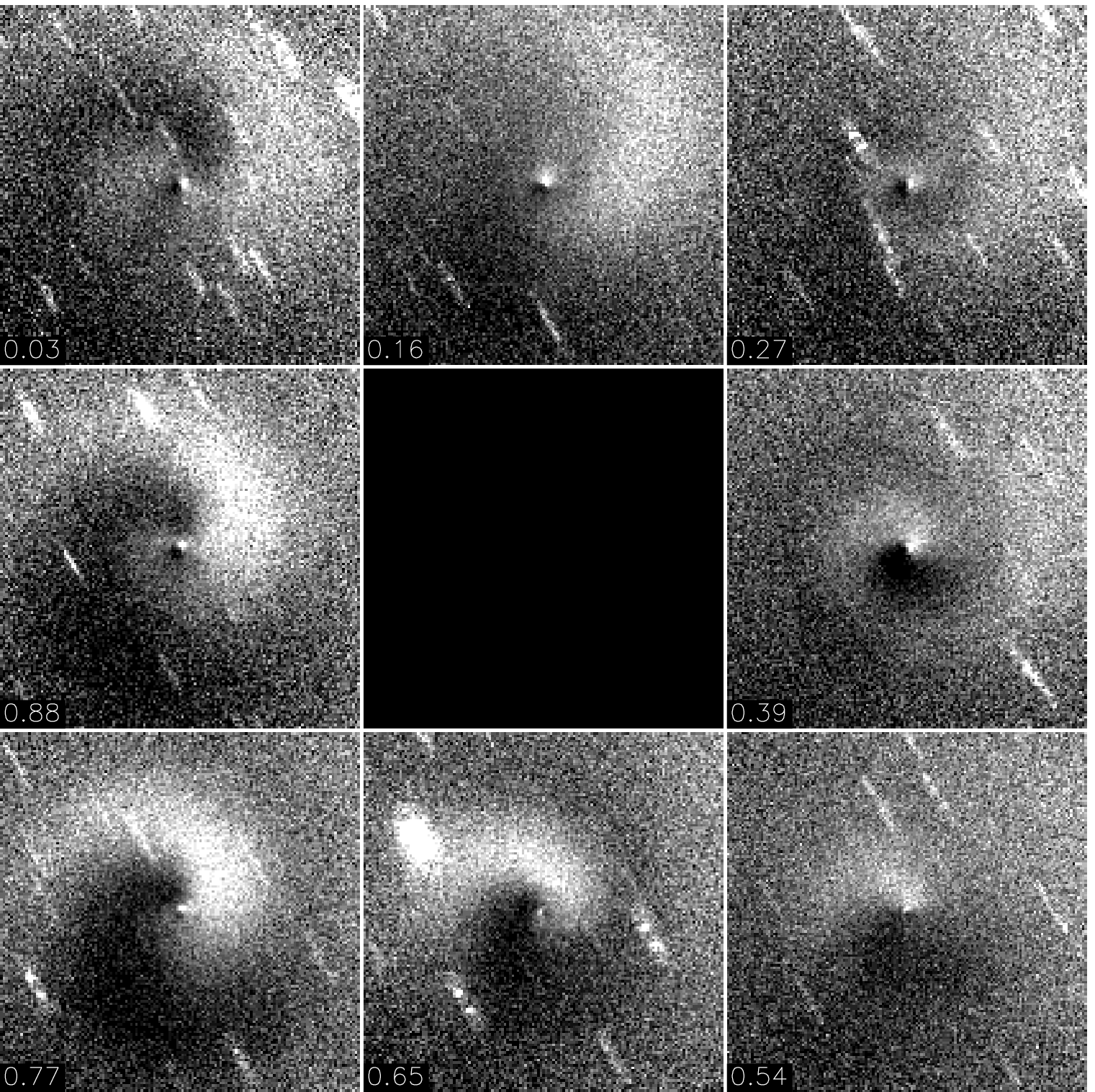}    
  \caption[Rotation cycle]{One complete rotation cycle (clockwise) constructed from images obtained on 2010 September 10--12. The rotational phase is given in the bottom left corner of each image, with zero phase set at 0:00 UT on September 9 and a rotation period of 17.2 hr. Each image is centered on the nucleus, has been enhanced by division of an azimuthal median profile, is approximately 50,000 km on a side at the comet, and is oriented so north is up and east is to the left. White represents the brightest areas and black represents the darkest areas. The images at 0.03, 0.65, 0.77, and 0.88 phase are from September 10, the images at 0.16, 0.27, and 0.39 phase are from September 11, and the image at 0.54 phase is from September 12.}
  \label{fig:rot_cycle}
\end{figure}

\begin{figure}
  \centering
  \epsscale{1.0}
  \plotone{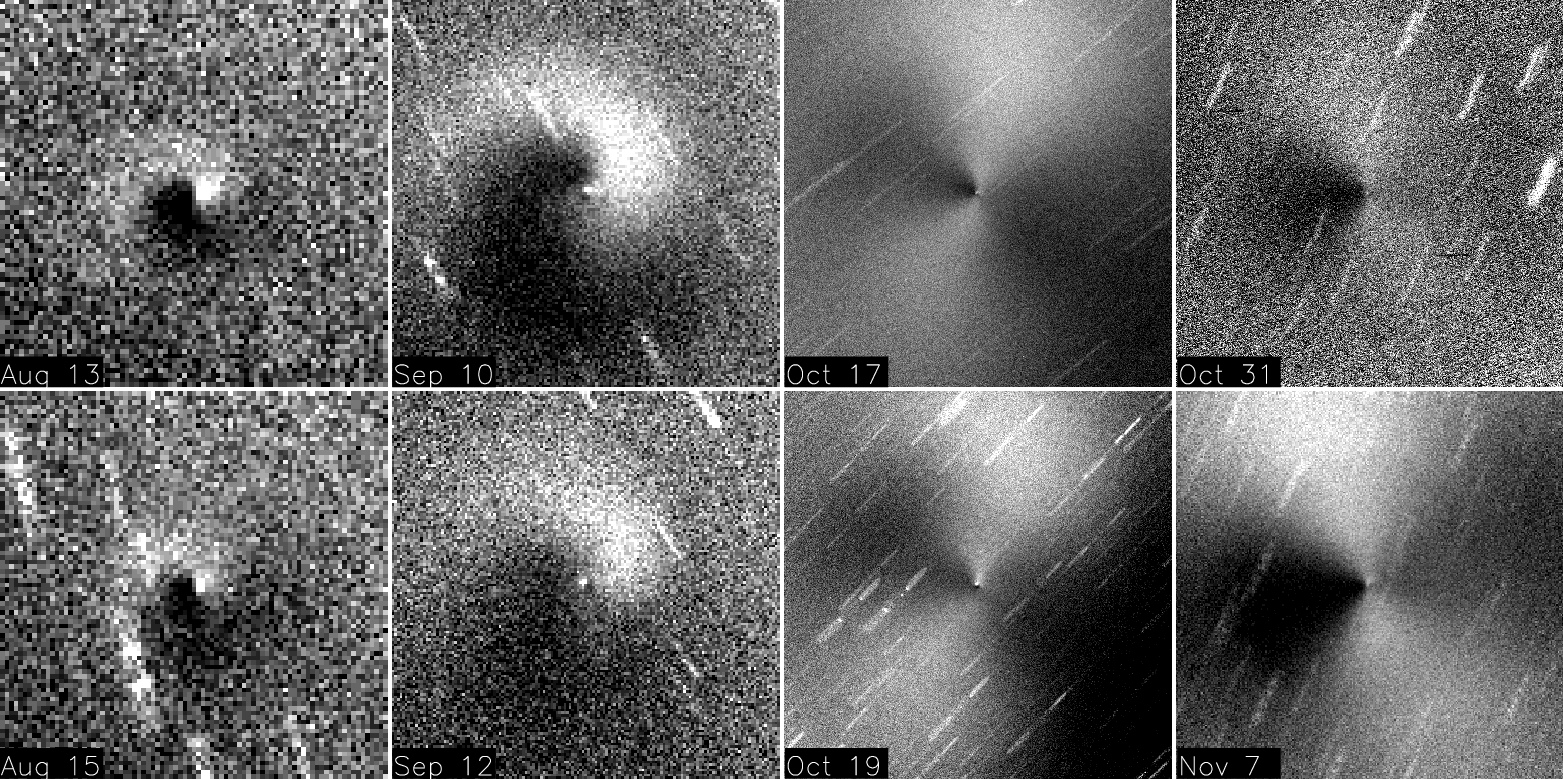}    
  \caption[Repeating morphology]{Representative pairs of images used to measure the rotation period. The first column shows images from August 13 and 15 (three cycles apart), the second column shows images from September 10 and 12 (three cycles apart), the third column shows images from October 17 and 19 (three cycles apart), and the fourth column shows images from October 31 and November 7 (nine cycles apart). Note that the image on October 31 is a snapshot observation from the 0.8-m with a relatively short exposure time while all other images are from the 1.1-m and have longer exposures. Each image is centered on the nucleus, has been enhanced by division of an azimuthal median profile, is approximately 50,000 km on a side at the comet, and is oriented so north is up and east is to the left. White represents the brightest areas and black represents the darkest areas.}
  \label{fig:same_phase}
\end{figure}

The bulk morphology repeated quasi-periodically, allowing the rotation period to be measured. Figure~\ref{fig:rot_cycle} shows an example of this, with a full rotational cycle constructed from images on September 10--12. Matching pairs of images were identified and the time interval between them was recorded (see Figure~\ref{fig:same_phase} for examples). The number of intervening rotational periods (``cycles'') was determined by comparing the morphology observed between the pairs and eliminating aliases. Throughout the apparition, significant rotational variation was observed during a given night without a repeating morphology. Depending on the temporal coverage (given in Figure~\ref{fig:phase_coverage}), this ruled out periods shorter than 6--9 hr. Periods longer than than this were ruled out either directly (i.e., images 24 hr apart looked very different) or by aliases (i.e., a 12 hr rotation period was ruled out because images 24 hr apart were different). Very long periods were not ruled out due to insufficient temporal coverage, but are unlikely due to the need for extreme solutions to explain the morphology repeating multiple times within the rotation cycle. 



\begin{figure}
  \centering
  \epsscale{1.0}
  \plotone{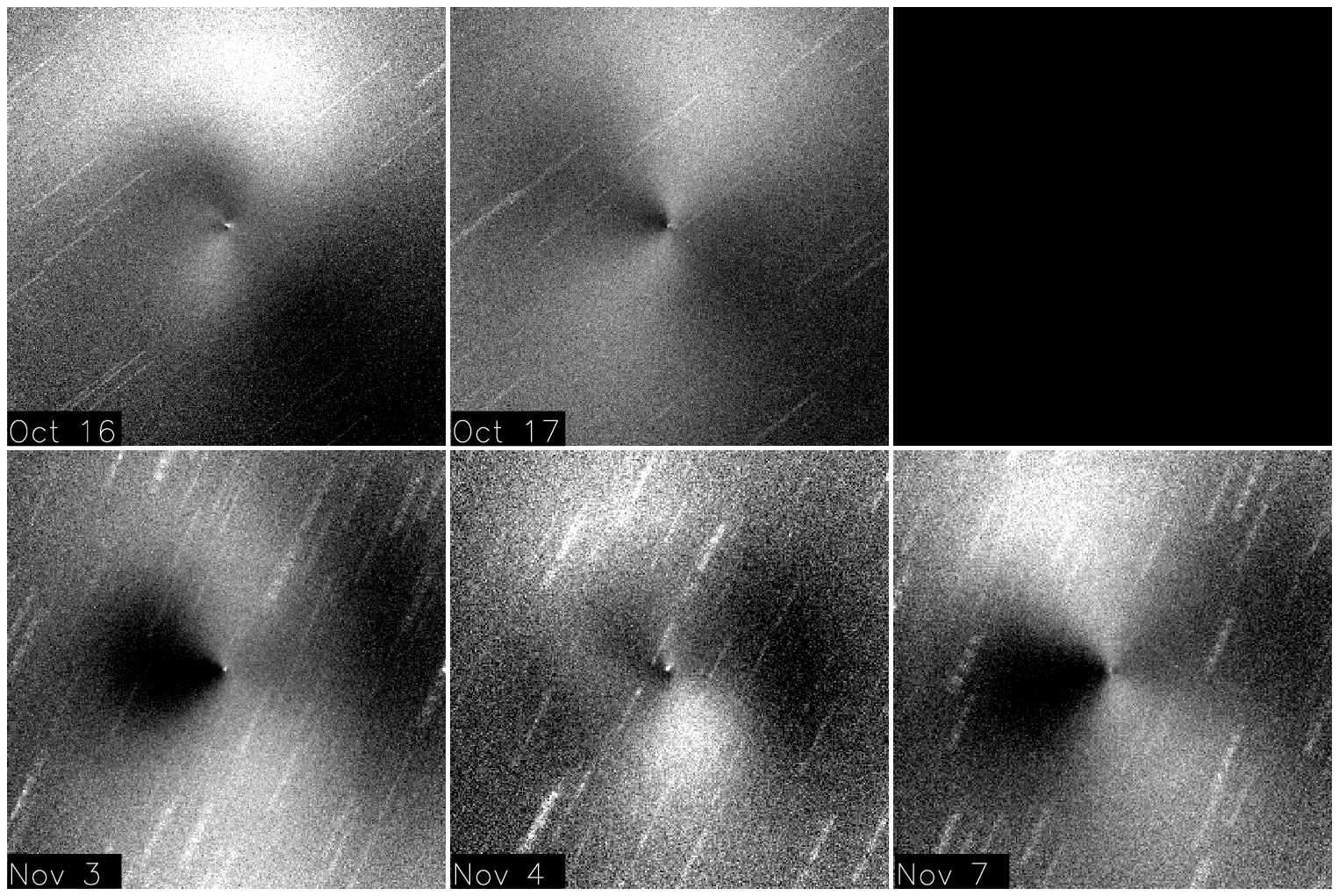}    
  \caption[Evidence for non-principal axis rotation]{Images at the same fundamental rotational phase for a given observing run but during different rotation cycles, showing evidence of non-principal axis rotation. Each image is centered on the nucleus, has been enhanced by division of an azimuthal median profile, is approximately 50,000 km on a side at the comet, and is oriented so north is up and east is to the left. White represents the brightest areas and black represents the darkest areas. The top row shows images taken one rotation cycle apart on October 16 and 17. 
The bottom row shows images taken at the same phase during three different cycles. The November 4 image is one cycle after the November 3 image while the November 7 image is five cycles after the November 3 image. Since the morphology best repeats after approximately three (and six) cycles, the November 7 image is equivalent to the morphology that would have been seen before the comet rose on November 5 (two cycles after November 3). See the text for further details of the differences between images. 
}
    \label{fig:complex}
\end{figure}

While image pairs could often be identified, the morphology did not repeat exactly each cycle. This can be seen by comparing images on consecutive nights in Figure~\ref{fig:cn_images}, and is shown explicitly in Figure~\ref{fig:complex} where we have selected images at the same rotational phase (based on the rotation periods we measure below) that show different morphology. The differences cannot be attributed solely to the changing viewing geometry, as the image from November 7 looks more similar to November 3 than does the image from November 4. While the general shape was often similar, the morphology within some images appeared to be curved while at other times it appeared more straight; also at times the morphology appeared to have rotated clockwise or counterclockwise in the plane of the sky. This suggests that the nucleus was not in simple, principal axis rotation, but has some non-principal axis rotation (i.e., a ``complex'' component). The magnitude of the complex component appears to be relatively small, as it was not obvious in our preliminary analysis of our August and September images, and only became problematic for determining a rotation period in October and November. We introduce the non-principal axis rotation here to familiarize the reader with the problems involved in determining the rotation period, and will discuss it in more detail in subsection~\ref{sec:complex_rot}.


While it became obvious later in the apparition that there was some complex component to the rotation, the dominant effect visible when looking at images is a general repetition of features. We call this the ``fundamental'' rotation period, with the complex component of the rotation period a secondary effect. Unless it is explicitly stated that we are referring to the complex component of the rotation period, subsequent mentions of a rotation period refer to the measured fundamental rotation period. Due to the complex component, image pairs one cycle apart were harder to match than image pairs multiple cycles apart and generally displayed much larger scatter in their derived rotation periods. As a result, when determining the rotation period for a given observing run, we have omitted rotation periods derived from a single cycle and instead have used the rotation periods measured over two or more cycles; note that the longer baselines also directly improve the accuracy of the results as compared to a single cycle. 

The fundamental rotation period was determined by dividing the time interval between matching images by the number of intervening cycles. All of the rotation periods were averaged to give the rotation period for a given run, with the date of this period corresponding to the average midpoint of all of the image pairs. The uncertainty in the rotation period was the standard deviation of the rotation period measurements. A summary of the matched pairs of images broken down by the number of intervening rotational cycles each month is given in Table~\ref{t:period_meas}. 
Because the morphology changed smoothly during a night, we can infer the morphology between images.
Thus, there were typically many additional pairs of images that do not quite match but help to constrain the rotation period by setting upper or lower limits.


\subsubsection{August}
We measured an average period of 16.67 $\pm$ 0.17 hr from eight matching pairs of images three cycles apart from August 13 to 17. There was a smaller change in morphology from cycle to cycle in August than in any other month, and the images one cycle apart suggest a similar rotation period.
The rotational coverage each night is plotted in the top panel of Figure~\ref{fig:phase_coverage}.


\subsubsection{September}
September provided the best temporal coverage of the apparition (second panel in Figure~\ref{fig:phase_coverage}). Even though we excluded September 9, when an extremely bright background star wiped out most of the coma morphology, we still identified 35 pairs of images from September 10 to 13. However, the complex component altered the morphology such that at some cycles the morphology repeated earlier while on other cycles it repeated later. Thus the rotation period measurements varied greatly for 1-cycle pairs (shown in Table~\ref{t:period_meas}). 
We found an average rotation period of 17.23 $\pm$ 0.13 hr from 22 pairs of images 3--4 cycles apart.

\subsubsection{October}
Poor weather reduced the temporal coverage and therefore the number of pairs of images identified in October (third panel in Figure~\ref{fig:phase_coverage}). Furthermore, unlike in August or September, the morphology looked different enough from cycle to cycle that matching pairs of images on consecutive nights was difficult. 
We identified 10 image pairs which were 3 or 5 cycles apart, resulting in a rotation period for October of 18.15 $\pm$ 0.15 hr.

\subsubsection{November}
In November the observing window shrank as the comet moved rapidly south and images one rotation cycle apart again looked somewhat dissimilar (fourth panel in Figure~\ref{fig:phase_coverage}). Thus, while images were obtained on eight consecutive nights and 12 out of 17 nights overall from October 31 until November 16, very few pairs of images were identified. The only matches were the three images from October 31 with six images on November 7 (two per image on October 31), yielding an average rotation period of 18.66 $\pm$ 0.07 hr.
However, there were a number of additional pairs where the rotational phase appeared to be slightly too early or late. Nearly all of these image pairs were a multiple of three cycles aparts (3, 6, 9, or 12 cycles) for a rotation period near 18.7 hr. Despite the worse signal to noise ratio on October 31 (images from the 0.8-m telescope with a shorter exposure time) and the low number of matching pairs, we consider the period to be accurate, but believe that our actual uncertainty for this time period was larger than the formal value of 0.07 hr. Thus we consider 18.7 $\pm$ 0.3 hr to be a reasonable estimate of the rotation period in early November.

\subsubsection{December}
Bad weather and instrument problems in December greatly limited our rotational coverage (fifth panel in Figure~\ref{fig:phase_coverage}). 
The first image of December 10 appears to be intermediate to the final two images of December 9, suggesting a rotation period near 19.0 hr. 

\subsubsection{January}
The comet faded rapidly between December and January making it difficult to discern morphology in the later month, and poor weather limited our rotational coverage (last panel in Figure~\ref{fig:phase_coverage}). Consequently, we cannot place any constraints on the rotation period.

\subsection{Changing Rotation Period}
\label{sec:changing_period}
\begin{figure}
  \centering
  \epsscale{1.0}
  \plotone{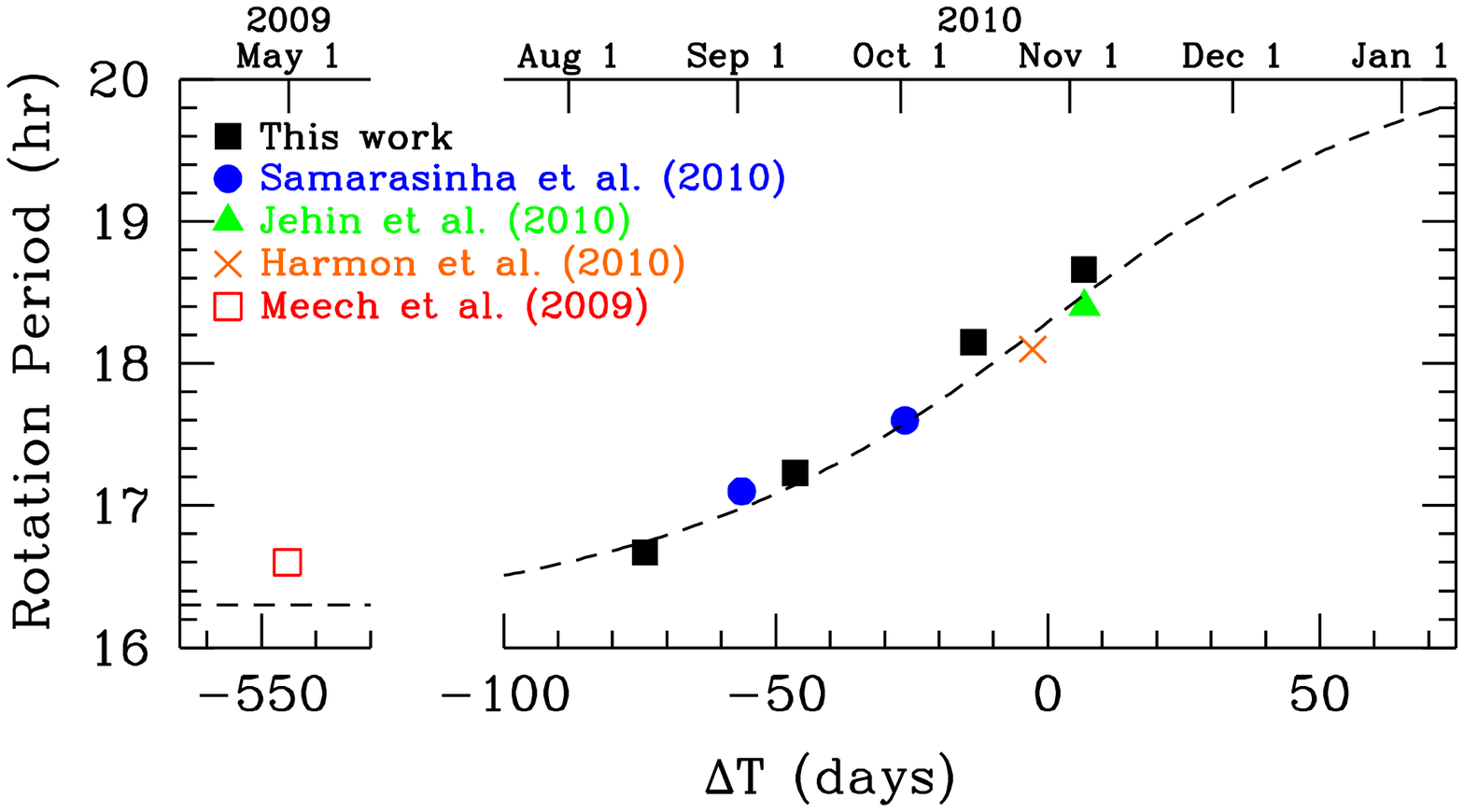}    
  \caption[Increasing rotation period]{The measured rotation period of Hartley 2 as a function of time relative to perihelion. Filled (black) squares are this work, filled (blue) circles are from \citet{iauc9178}, the filled (green) triangle is from \citet{cbet2589}, the (orange) cross is from \citet{iauc9179a}, and the open (red) square is from \citet{meech09}. Error bars were not given by all authors, but are likely 0.2--0.4 hr. The times of our measurements were set as the average midpoint of our image pairs (given in Table~\ref{t:period_meas}) while the time of the measurements by other authors was set as the midpoint of the range over which the rotation period was measured. The midpoint for \citet{meech09} was set to 2009 May 1 since no date range was provided in the abstract. The dashed curve is an arbitrary  function in which the nucleus experiences torquing $\propto$ r$_H$$^{-3}$ when r$_H$ $<$ 2 AU and no torquing for r$_H$ $\ge$ 2 AU. The function has been scaled to roughly agree with the plotted points as well as our estimate of near 19.0 hr for December 9--10 and the report by \citet{cbet2589} noting a change from about 18.2 to 19 hours from October 29 until December 7.
}
  \label{fig:period_change}
\end{figure}

Our data indicate that the rotation period increased steadily from August through November (and possibly December). This increase is consistent with the rotation periods reported by other researchers: near 17.1 hr from September 1 to 3 \citep{iauc9178}, around 17.6 hr from September 30 to October 4 \citep{iauc9178}, 18.1 $\pm$ 0.3 hr from October 24 to 27 \citep{iauc9179a}, and 18.4 $\pm$ 0.3 hr from October 29 to November 12 \citep{cbet2589}. We plot each of these reported rotation periods along with our own periods for August through November in Figure~\ref{fig:period_change}, with the midpoint of each dataset used as the date of the measurement. The increase in rotation period is roughly linear, with a slope of $\sim$0.02 hr day$^{-1}$ during the 80 days between our August and November data. This matches the results reported by \citet{cbet2589} whose observations from October 29 until December 7 implied an increase in the rotation period corresponding to 2 hr in 100 days.


There is no reason to expect the change in rotation period to be linear. If the change is driven by asymmetric torquing of the nucleus due to outgassing, the effect should be larger at smaller heliocentric distances. Furthermore, the measurement by \citet{meech09} of a rotation period about 16.6 hr near aphelion demonstrates that very little torquing had occured between 2009 May (r$_H$ $\sim$ 4.5 AU) and 2010 August (r$_H$ $\sim$ 1.5 AU). We attempted to fit all of the rotation period measurements with a single function, finding a satisfactory fit for torquing $\propto$r$_H$$^{-3}$ when r$_H$ $<$ 2.0 AU and no torquing when r$_H$ $\ge$ 2.0 AU (plotted in Figure~\ref{fig:period_change} as a dashed line). Lower order functions were too shallow while higher order functions implied a significantly longer rotation period in December than appears to be the case based on our own estimate of near 19.0 hr and the results reported by \citet{cbet2589} of a change from about 18.2 to 19 hours from October 29 until December 7. While this function was determined empirically, its slope is similar to the water sublimation rate given by the model of A'Hearn\footnote{http://www.astro.umd.edu/$\sim$ma/evap/index.shtml} based on the method described by \citet{cowan79}. Taking the average slope of the water sublimation over the heliocentric range of our Hartley 2 observations (r$_H$ = 1.06--1.46 AU) and assuming the comet is a rapid rotator, we find a water sublimation rate of r$_H$$^{-2.9}$ for a rotation axis perpendicular to the comet-Sun line (consistent with our preferred solution which will be discussed in subsection~\ref{sec:modeling}) or r$_H$$^{-2.4}$ for a rotation axis pole-on to the comet-Sun line. Furthermore, the truncation in the torquing at 2.0 AU in our function is a crude approximation of the severe decrease in efficiency of sublimation which occurs around 2.0--2.5 AU in the model as insolation increasingly goes into heating the ice rather than into sublimation.

Assuming that the torquing is due to a physical process that depends on the heliocentric distance, it should be symmetric about perihelion (although there may also be a seasonal effect that is not symmetric). The rotation period increased from $\sim$16.6 hr to $\sim$18.3 hr on the inbound leg of the orbit, so if a similar spin-down occured on the outbound leg, the total spin-down for the whole orbit should be $\sim$3.4 hr. We are unaware of any rotation period measurements later than early 2010 December (our own observations from December 9--10 and the observations by \citet{cbet2589} which ran through December 7), but based on our crude fit, we predict that the rotation period should have been near 19.8 hr in 2011 January and should be near 20.3 hr when Hartley 2 reaches aphelion.

It is possible that the nucleus has spun-down by similar amounts each perihelion passage. Hypothetically if the spin-down had been occurring indefinitely, at some point in the recent past the nucleus would have been spinning so fast that it would have disrupted. Following the method of \citet{mueller96} as adapted from \citet{luu92} we can determine the critical period P at which a nucleus with no internal strength will split, 

\begin{equation}
P = \sqrt{\frac{-2\pi}{GF(f)\rho}}
\end{equation}

\noindent where G = 6.67$\times$10$^{-11}$ N kg$^{-2}$ is the gravitational constant, $\rho$ is the density of the spheroid, and F($f$) is a function of the axis ratio $f = b/a$ (defined in Equation 9 of \citet{luu92}). Assuming a density of 400 kg m$^{-3}$ and an axis ratio of 0.5, this yields a critical rotation period of 7.2 hr. This should be considered an upper limit on the critical rotation period as any internal strength the comet has will allow it to spin faster before disrupting. 

If the spin-down observed this apparition has occurred at a comparable level on previous orbits, the rotation period would have been faster than the critical rotation period three orbits ago. However, if the torquing remained constant, the spin-down on previous orbits would have been smaller than on the current orbit due to the comet's larger angular momentum in the past. Accounting for this, the rotation period still would have been faster than the critical rotation period about a dozen orbits ago. Since this scenario implies that the comet could not have existed in the recent past, this rate of torquing cannot have been occurring long term. We suggest three possible explanations: 

\begin{enumerate}
\item The comet was previously spinning-up, reached the critical rotation period and broke up, and the remaining piece is now spinning down. We do not have any evidence that Hartley 2 broke up in the recent past, however, it was only discovered in 1986 and it is possible that smaller fragments rapidly disappeared after a break up, leaving the current nucleus as the only surviving remnant. A newly exposed active region could then cause the necessary torquing to begin slowing the rotation down.

\item This magnitude of torquing is a recent phenomenon and previous spin-down was smaller or non-existent. New or larger torquing than was previously occurring could be due to a freshly exposed active region. This could be caused by many scenarios such as fragmentation, impact, build up of pressure below the surface causing an explosive outburst, or a change in the orientation of the nucleus to expose a previously shadowed region. Alternatively, some areas could have become less active or completely inactive with the effect being to increase the relative torquing caused by the regions which remained active. 

\item The complex component interacts with the fundamental period in such a way as to cause the fundamental period to appear to be spinning-down when it is not (or the real spin-down is much smaller). A demonstration of this would require a very high level of knowledge about the rotational state and is likely unresolvable from any single data set. However, the combination of the EPOXI spacecraft data with numerous ground-based datasets might allow a good enough determination of the rotation to test this scenario.
\end{enumerate}

Our unpublished photometry shows that there is a secular decrease in the gas production rate since the 1991 and 1997 apparitions. This may be affecting the rotation period, but until a comprehensive model of the nucleus is obtained, the effect of a secular decrease in the production rates is unknown. Further study, particularly nucleus lightcurve measurements when Hartley 2 is near aphelion (in 2013--2014), will reveal if the rotation period continues to evolve and may help determine what is causing the spin-down. 


\subsection{Non-Principal Axis Rotation}
\label{sec:complex_rot}
As introduced briefly in subsection~\ref{sec:rot_period}, the morphology did not repeat exactly each cycle, which we interpret as being due to a small non-principal axis rotation. This is shown in Figure~\ref{fig:complex} for images in October and November. These images demonstrate the key manifestations of the complex component interacting with the fundamental rotation: the morphology appears to be curved at some times and straighter at other times, and appears to rotate in the plane of the sky.
The images on October 16 and 17 are exactly one cycle apart, but the northern feature on October 16 is curved while the same feature on October 17 is straighter. 
There is also a faint feature extending from the nucleus towards the northeast on October 16 which is not evident on October 17. The November 3 and 7 images are approximately linear while the November 4 image is curved.
The orientation of the two features is nearly north-south on November 3 while it is rotated to the north-northeast on November 4 and 7. Finally, the November 4 image has a faint feature connecting the northern and southern features to the east which appears to be absent on November 3 and 7.

The morphological difference from cycle to cycle was minimal in August. In September the morphology appeared relatively similar but at times it repeated earlier or later than expected. These differences appear to be due to the interaction of the complex component with the fundamental rotation period, with the effect of the average interaction being relatively benign. In October and November the morphology looked quite different from cycle to cycle, making it difficult to match images one rotation cycle apart. At times the complex component worked in tandem with the fundamental rotation to create stronger curvature while at other times the complex componenet appeared to nearly cancel out the fundamental rotation, causing the jets to appear nearly straight. We had insufficient temporal coverage in December and January to observe the effects of the complex component. We interpret the varying evidence of the complex component during the apparition as being due to the changing viewing geometry, as we saw the complex component occurring from different viewpoints during the apparition. 
Our modeling (subsection~\ref{sec:modeling}) suggests that throughout the apparition the total angular momentum vector (our average PAs) is within $\sim$15--25$^\circ$ of the instantaneous fundamental rotation axis, thus the motion of the complex component is only $\sim$15--25$^\circ$.

Throughout the apparition, the best rotational matches were found for images multiples of three cycles apart. This suggests that the period of the complex component was close to three times the fundamental rotation period. Since we measured fundamental rotation periods from 16.7--18.7 hr during the apparition, this implies that the complex period was 50--56 hr. When images were phased to triple the fundamental rotation period for a given observing run, the November images were in better agreement than the September images, suggesting that the complex rotation period is closer to 56 hr than to 52 hr (assuming that it is unchanging).

At this time we cannot determine the complex component period more precisely or determine if it is increasing (or decreasing) with the increasing fundamental rotation period. With sufficient temporal coverage, the morphology at integer numbers of fundamental cycles could be compared to restrict the complex period. That is, if the fundamental period was 18 hr and the complex period was 54 hr, morphology should exactly repeat every three fundamental cycles (ignoring the changing viewing geometry). However, if the fundamental period was 18 hr and the complex period was 51 hr, the end of the third fundamental cycle would occur 3 hr after the complex cycle repeated itself. This difference would compound after six cycles, making the morphology a better match after three cycles than six (noting that the viewing geometry would have also changed more after six cycles). With enough temporal coverage, small differences between integer numbers of cycles could be identified and the complex period restricted. In theory, modeling of the different morphology each cycle could also constrain the complex rotation period, but the number of free parameters is too large for this to be reasonable, and our temporal coverage is insufficient. Until the value of the complex period is further constrained, we cannot investigate whether or not it is increasing along with the fundamental rotation period. We hope that combining our data with those of other observers might lead to a more robust understanding of the non-principal axis rotation of the nucleus of Hartley 2.

Depending on the specific nature of the non-principal axis rotation, what we have identified as the fundamental rotation period might be a beat frequency of the true rotation period and the complex component. A simple example of this involves the fundamental and complex rotations occurring in the same direction, causing the morphology to repeat faster than either rotation period. 
At this time we do not understand the rotation state well enough to investigate this possibility, but simply note that the fundamental rotation periods we measured may not directly correspond to the rotation rate of the nucleus.

\subsection{Key Constraints on a Nucleus Rotation and Coma Jet Model}
\label{sec:modeling_constraints}
While our data are insufficient to determine a unique solution, they yield several strong constraints on model parameters.
We list below the key morphological characteristics in approximate decending order of importance.
\begin{itemize}
\item November has features $\sim$180$^\circ$ apart that each appear to be side-on corkscrews. There is no evidence of a face-on spiral from either feature, implying that the Earth was outside of both swept out cones of material. We conclude that the view was more equator-on at this time than at any other time during the apparition. August and September are dominated by a northern feature that appears as a nearly complete spiral indicating that we are within the cone of material (i.e., the source region is pointing either towards or away from us). Images taken during December and January are too faint and sparsely sampled to get a full rotation cycle, but the morphology suggests we are within the cone of material again. 
\item The average PA of the northern feature during November was 10$^\circ$ $\pm$ 5$^\circ$, which we consider to be the best determined PA of the apparition. Additional, less certain, PAs for the northern feature were measured for August (350$^\circ$ $\pm$ 20$^\circ$), September (335$^\circ$ $\pm$ 20$^\circ$), October (350$^\circ$ $\pm$ 10$^\circ$), and December (55$^\circ$ $\pm$ 15$^\circ$). PAs for the southern feature were measured for October (145$^\circ$ $\pm$ 10$^\circ$), November (190$^\circ$ $\pm$ 5$^\circ$), and December (215$^\circ$ $\pm$ 15$^\circ$).
\item Overall, the migration of the northern feature from the north-northwest in August and September to the northeast in December and January is strongly suggestive that this is the total angular momentum vector changing its projected position angle with viewing geometry.  A caveat: there appears to be more day/night varition in intensity of the northern feature late in the apparition than early on. Since the Sun is also towards the east, the average center of brightness of the feature is likely eastward of the projected pole.
\item In August and September the spiral moves clockwise, while in December and January it moves counterclockwise. This can be explained by the changing viewing geometry, as the Earth was looking at it from a direction $\sim$120$^\circ$ different in December and January than in August and September. It is difficult to determine the sense of rotation in October and November due to the side-on morphology causing the features to overlap.
\item There was a bulk change in relative brightness throughout the apparition. The north was clearly brighter in August and September, however, later in the apparition the brightnesses were roughly comparable (October through December) or possibly brighter in the south (January, although we do not have complete rotational coverage). This suggests that the sub-solar point moved south during the apparition.
\item In October there is a clear second source to the south for the first time, although it may be there in August and September. The two jets appear to overlap towards the east at some rotational phases in October and November. 
\item The angular extent of the swept out regions in November (when the sub-Earth point is nearly at the equator) constrains how close to the pole the source regions are located.
\item The interaction of the fundamental rotation with the complex component of the rotation yields some morphology that is almost linear while at other times it appears curved.
\end{itemize}

\subsection{Model Practicalities}
\label{sec:modeling_background}
Creating a model to reproduce a given comet's coma morphology is always a challenge for a variety of reasons, but primarily because the underlying 3-D structures are always viewed projected on the plane of the sky. Any single observed morphology can usually be reproduced by a family of 3-D solutions, of which only one (if any) correctly represents the comet. Multiple observations at differing rotational phases and/or differing viewing geometries each yield additional solution sets but, hopefully, also constrain the solution due to the requirement that the correct solution must match all of the images. In practice, numerous assumptions and simplifications must be made to create a practical numerical Monte Carlo model, such as the number, location and shape of source regions, outflow velocities, dispersion characteristics of the particles, and the response to solar insolation. We have developed a model over the past dozen years (see \citet{schleicher03a}) that uses many of these properties, along with the orientation of the rotation axis, the rotation period, and several others.

With our discovery of a discrete CN feature in August \citep{iauc9163a}, we anticipated that the combination of rotational measurements early and late in the apparition -- when the viewing geometry was changing very slowly -- coupled with a rapid change in geometry near perigee, would readily yield a determination of the pole orientation and the location(s) of the source region(s) on the nucleus based on the assumption that the feature was directly associated with a jet of material released from isolated source region(s). The subsequent evidence for complex rotation (subsection~\ref{sec:complex_rot}), further complicated by an apparently rapidly increasing fundamental rotation period (subsection~\ref{sec:changing_period}), has instead resulted in a much more difficult problem. In particular, even with our extensive temporal coverage, there remain significant gaps within any 3-cycle (50--56 hr) phasing associated with a given observing run, and it is these gaps that make it very difficult to disentangle the fundamental and secondary complex rotation components. Therefore the modeling results presented next and based on our data alone must be considered very preliminary, self-consistent but not unique; we expect that subsequent intercombining of data from other observers will further constrain models and hopefully yield the correct and definitive answer.

\subsection{Preliminary Modeling of the CN Morphology}
\label{sec:modeling}
As noted previously in Section~\ref{sec:modeling_constraints}, one of the apparent strongest constraints on the nominal rotational axis (i.e. the total angular momentum vector) is the side-on, corkscrew-like appearance of both the northern and southern features in early November. Not only do these features provide the best determined position angle of the axis throughout the apparition, yielding a great circle of deprojected pole solutions, but there is also the least amount of overlap between the two features, directly implying that we are observing from close to the comet's equator. These measurements directly yield two solution regions 180$^\circ$ apart. Only one of these, however, yields the clockwise rotation observed in August and September. This solution region is centered at an RA of 257$^\circ$ and Dec of $+$67$^\circ$, based on an obliquity of the ``axis'' to the normal of the comet's orbital plane of 10$^\circ$ and a principal angle along the orbit of 250$^\circ$. Based on the scenario described next and when fitting of the features became problematic, we estimate that this solution region has a radius (or uncertainty) of about 15$^\circ$.

\begin{figure}
  \centering
  \epsscale{1.0}
  \plotone{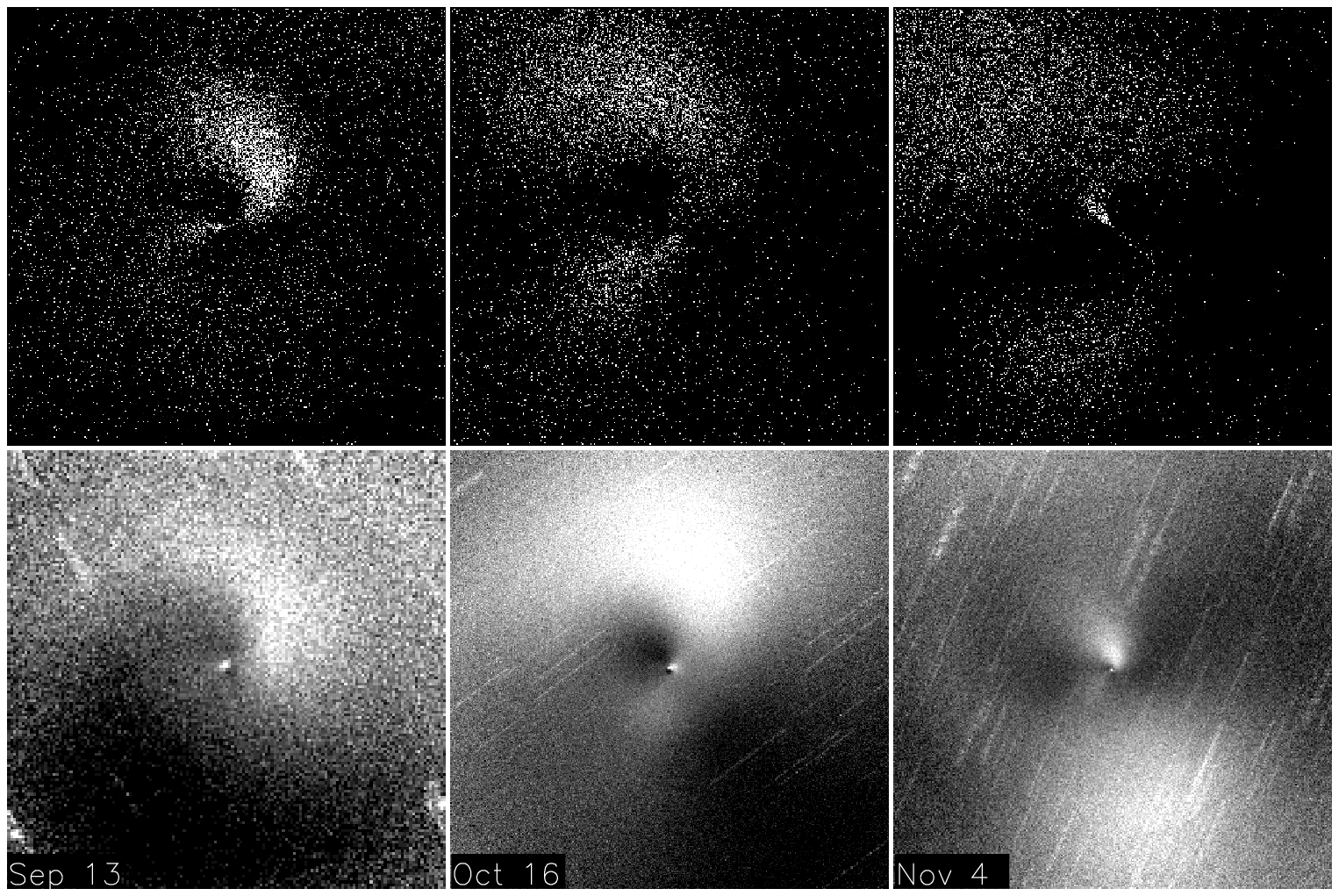}    
  \caption[Model images]{Model (top row) with corresponding images (bottom row). The date of each image pair is given in the bottom left corner of the lower image. Each image is centered on the nucleus, is approximately 50,000 km on a side at the comet, and is oriented so north is up and east is to the left. Model images were enhanced by removal of a $\rho$$^{-1}$ profile while data images were enhanced by division of an azimuthal median profile. While these enhancements are different, for this preliminary model they are sufficient to remove the bulk fall off and reveal the underlying morphology. White represents the brightest areas and black represents the darkest areas. The model solutions shown here do not include a complex component so these images were chosen because the morphology appeared least affected by the complex rotation. See the text for further details.}
  \label{fig:model_images}
\end{figure}

Using this solution region for the total angular momentum vector, we were next able to constrain the two source locations on the surface of the nucleus. The apparent shape and orientation of the features as functions of both rotation and viewing geometry throughout the apparition (see subsection~\ref{sec:morph_evolution}) indicated that each jet originates from a mid-latitude source, one in each hemisphere. The least well-determined components of the solution are the relative longitudes of the sources, which we tentatively place about 120$^\circ$ apart. Because our solution is very preliminary, we have not attempted to fine-tune a variety of other parameters, such as outflow and dissociation velocities, directional and velocity dispersions, and the variation of outgassing with Sun angle. Even so, our solution reproduces the overall morphology remarkably well, as evidenced in Figure~\ref{fig:model_images}, where we show a representative image from each of the September, October, and November runs, along with this first-cut model for each.


In September (left-hand pair in Figure~\ref{fig:model_images}), the CN morphology appeared as a nearly complete spiral viewed from the interior of the swept-out cone from a single jet tilted towards the northwest. A mid-latitude primary source produces the dominant feature in the north and west. However, our model, having a sub-Earth latitude at this time of $-$40$^\circ$, suggests that the southeastern portion of the observed feature is influenced and perhaps dominated by the overlap of a southern jet, and that we are not as interior to the swept-out cone as first thought. In this scenario, the southern source is active throughout the apparition, but being smaller in size, its jet usually appears fainter than the northern jet. In mid-October (center pair), the two jets were more separated as the sub-Earth latitude approached the equator ($-$26$^\circ$). By the night of the EPOXI encounter (right-hand pair), the sub-Earth latitude had crossed the equator and is $\sim$$+$12$^\circ$, for a near side-on view, but the Sun had a PA of 103$^\circ$ and the phase angle ($\theta$) is 59$^\circ$, putting the majority of the gas released in the daytime hours towards the east. Each of the jets appeared as broad, partially complete corkscrews, towards the north and south.

Note that since this model solution does not include a complex rotational component, we have intentionally attempted to fit images least affected by this second order effect. For instance, as shown in Figure~\ref{fig:complex}, some rotational cycles show stronger curvature when the complex motion is working in tandem with the rotational effects, while at other times the jets appear nearly straight when the complex motion is nearly cancelling out the rotational curvature; we have chosen the intermediate case to fit. Also note that there is still complex motion ongoing in these images, only their appearances are less affected due to the interaction of the complex and fundamental rotations. By adding simple precession (with a period of $\sim$3 times that of the fundamental rotation), we have briefly explored the effects of complex rotation on the observed morphology, confirming that these bulk complex attributes can indeed be reproduced. But, until the specific dynamical state of the nucleus is ascertained (i.e. rotation and precession and/or rolling and/or nodding, etc.), it is not realistic to attempt to create a more complete model.

\subsection{Hartley 2 at the Time of the EPOXI Flyby}
\begin{figure}
  \centering
  \epsscale{0.7}
  \plotone{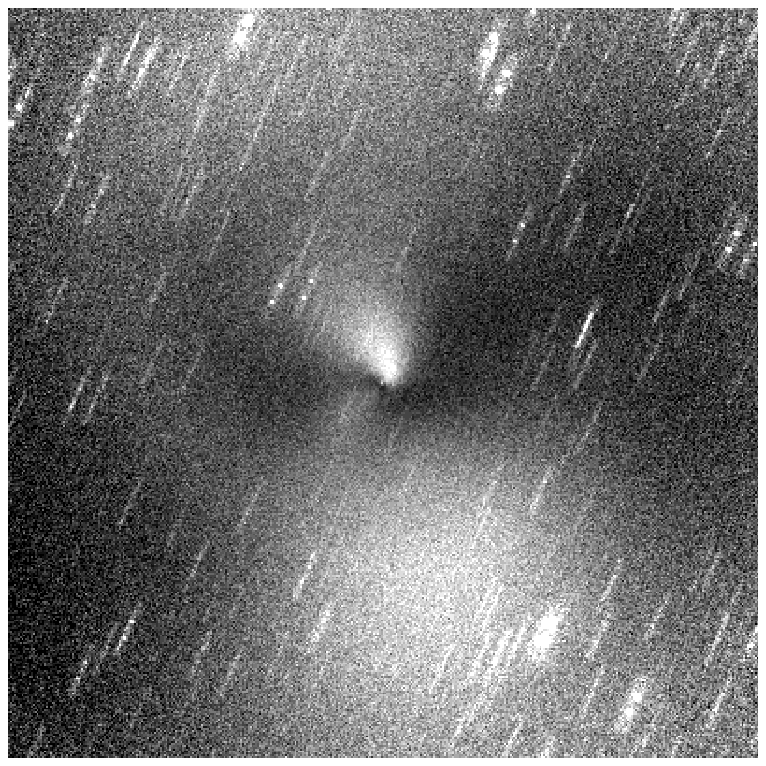}    
  \caption[Image close to time of flyby]{CN image of Hartley 2 shortly before the EPOXI flyby. This image was acquired at 12:11 UT on 2010 November 4, less than 2 hr prior to the encounter. 
The image is centered on the nucleus, has been enhanced by division of an azimuthal median profile, is approximately 50,000 km on a side at the comet, and is oriented so north is up and east is to the left. White represents the brightest areas and black represents the darkest areas. The morphology at flyby should have looked relatively similar to that shown here, but with the diffuse blob to the south having moved slightly further from the nucleus and the feature to the north having grown brighter and expanded  towards the north-northeast. We acquired an additional image during twilight at 12:59 UT that has much lower signal to noise but confirms this morphology.}
  \label{fig:flyby}
\end{figure}

Combining what we have learned in subsections~\ref{sec:morphology}--\ref{sec:modeling}, we can assess the state of Hartley 2 at the time of the EPOXI flyby on 2011 November 4 at 14:01 UT. Figure~\ref{fig:flyby} shows the morphology at 12:11 UT, less than 2 hr before the flyby. We acquired an additional image during twilight at 12:59 UT; although the signal to noise was considerably worse it shows very similar morphology. Extrapolating from this image to the time of the flyby, the broad southern feature and the very faint northern feature at the top of the image would have moved slightly farther from the nucleus, and the bright feature extending from the nucleus to the north-northeast would have brightened, expanded, and extended slightly further north-northeast. The fundamental rotation period at this time was near 18.7 hr. Based on our preliminary Monte Carlo jet solution and ignoring the complex component we predict that the sub-Earth point was at $+$12$^\circ$ and the sub-solar point was at $+$4$^\circ$. Knowledge of what the comet was doing during the encounter is important for giving context to the ``ground truth'' observations made by the spacecraft. It is hoped that as more complete models of the nucleus' rotation state are constructed, the large scale morphology we observed from the ground can be linked to the near-nucleus jets and specific active regions on the surface observed by EPOXI.


\section{Summary}
\label{sec:summary}
We observed Hartley 2 at Lowell Observatory for 39 nights between 2010 July and 2011 January. Images were obtained with broadband R and narrowband comet filters, however, this paper is restricted to the narrowband CN images as they best revealed coma morphology. We summarize our key results below.

\begin{itemize}
\item {\it CN morphology --} No morphology in CN emission was observed in 2010 July; we discovered CN morphology in 2010 August and continued to see morphology through 2011 January. The morphology changed smoothly during a night, and repetition of the morphology on different nights allowed us to measure the rotation period.
\item {\it Evolution of CN morphology --} A CN feature was seen in August generally towards the north. September's morphology looked similar to August's but showed hints of a possible second CN feature towards the southeast. From October through December two CN features were distinctly visible, while only the southern feature was evident in January. The overall orientation of the morphology gradually moved counterclockwise from the northwest to north to northeast during the apparition. 
\item {\it Evidence for non-principal axis rotation --} The general morphology repeated from night to night, although differences were observed on consecutive rotation cycles. This suggests a small non-principal axis rotation. The morphology best repeated after three cycles, implying the complex component had a period of $\sim$3$\times$ the fundamental rotation period.
\item {\it Rotation period --} We measured rotation periods of 16.67 $\pm$ 0.17 hr from August 13--17, 17.23 $\pm$ 0.13 hr from September 10--13, 18.15 $\pm$ 0.15 hr from October 12--19, and 18.7 $\pm$ 0.3 hr from October 31--November 7. Further, we estimated a rotation period near 19.0 hr from December 9--10. Combining these measurements with other published rotation periods, we find that Hartley 2 spun down during the 2010 apparition. The rate of spin-down is consistent with torquing that varies as r$_{H}^{-3}$. A spin-down of the magnitude observed from 2010 August through 2010 November cannot have occurred for more than a dozen orbits or the nucleus would have been spinning so fast in the past that it would have disrupted.
\item {\it Monte Carlo Jet Modeling --} We used a number of characteristics of the morphology to guide numerical modeling of the pole orientation and locations of source regions on the nucleus. Our preferred, self-consistent but non-unique pole solution has an obliquity of 10$^\circ$ relative to the normal of the comet's orbital plane (corresponding to a pole oriented towards RA = 257$^\circ$ and Dec = $+$67$^\circ$) and two mid-latitude sources, one in each hemisphere.
\end{itemize}

This work gives a preliminary understanding of the rotation period, non-principal axis rotation, pole orientation, and locations of active regions, but we anticipate that these quantities will be far better constrained in the near future. Due to the unique circumstances of Hartley 2's 2010 apparition, there are a wealth of datasets which, when combined, may yield unprecedented insight into the rotation period(s), pole orientation, and number and location(s) of active regions on the nucleus. 
Analyses of many of these datasets are underway, 
and we look forward to gaining a richer understanding of Comet Hartley 2.

\section*{Acknowledgements}
We thank the anonymous referee for a prompt and helpful review. Thanks to Brian Skiff and Len Bright for sacrificing some observing time and helping us acquire snapshot images in between our scheduled observing runs. Thanks to Larry Wasserman for also giving up some observing time and for generating the robotic scripts for the 0.8-m telescope. Thanks to Eddie Schwieterman for helping observe with the 1.1-m telescope in July and August. We are grateful for JPL's Horizons for generating observing geometries. This work was supported by NASA planetary astronomy grant NNX09B51G.



\clearpage

\renewcommand{\arraystretch}{0.6}
\renewcommand{\baselinestretch}{1.63}

\begin{deluxetable}{lccrccccc}  
\tabletypesize{\scriptsize}
\tablecolumns{10}
\tablewidth{0pt} 
\setlength{\tabcolsep}{0.05in}
\tablecaption{Summary of Hartley 2 observations and geometric parameters\,\tablenotemark{a}.}
\tablehead{   
  \colhead{UT}&
  \colhead{UT}&
  \colhead{Telescope}&
  \colhead{$\Delta$T}&
  \colhead{$r_H$}&
  \colhead{$\Delta$}&
  \colhead{$\theta$\tablenotemark{b}}&
  \colhead{PA Sun\tablenotemark{c}}&
  \colhead{Conditions}\\
  \colhead{Date}&
  \colhead{Range}&
  \colhead{Diam. (m)}&
  \colhead{(days)}&
  \colhead{(AU)}&
  \colhead{(AU)}&
  \colhead{($^\circ$)}&
  \colhead{($^\circ$)}&
  \colhead{}
}
\startdata
2010 Jul 19&10:00--10:22&1.1&$-$100.8&1.68&0.86&29&44&Thin cirrus\\
2010 Aug 13&03:28--11:48&1.1&$-$75.9&1.46&0.57&30&22&Photometric\\
2010 Aug 14&03:03--11:48&1.1&$-$74.9&1.45&0.56&30&21&Photometric\\
2010 Aug 15&03:15--12:10&1.1&$-$73.9&1.45&0.55&30&20&Clouds\\
2010 Aug 16&07:40--11:54&1.1&$-$72.8&1.44&0.54&31&19&Clouds\\
2010 Aug 17&07:48--11:01&1.1&$-$71.9&1.43&0.53&31&18&Clouds\\
2010 Sep 9&02:53--12:16&1.1&$-$48.9&1.25&0.33&37&357&Photometric\\
2010 Sep 10&02:36--12:09&1.1&$-$48.0&1.24&0.32&37&357&Photometric\\
2010 Sep 11&02:33--12:08&1.1&$-$47.0&1.24&0.31&37&356&Photometric\\
2010 Sep 12&02:33--12:11&1.1&$-$46.0&1.23&0.31&38&356&Clouds\\
2010 Sep 13&02:35--12:11&1.1&$-$44.9&1.22&0.30&38&355&Clouds\\
2010 Oct 12&03:12--12:34&0.8&$-$15.9&1.08&0.13&49&43&Photometric\\
2010 Oct 13&03:13--12:43&0.8&$-$14.9&1.08&0.13&49&48&Photometric\\
2010 Oct 14&03:13--12:40&0.8&$-$13.9&1.08&0.13&50&52&Photometric\\
2010 Oct 15&03:06--05:33&0.8&$-$13.1&1.07&0.13&50&56&Clouds\\
2010 Oct 16&05:01--12:21&1.1&$-$11.9&1.07&0.12&51&61&Thin cirrus\\
2010 Oct 17&05:00--12:38&1.1&$-$10.9&1.07&0.12&51&65&Clouds\\
2010 Oct 19&10:56--12:24&1.1&$-$8.8&1.07&0.12&52&73&Clouds\\
2010 Oct 31&07:10--12:36&0.8&$+$3.2&1.06&0.14&58&99&Thin cirrus\\
2010 Nov 1&07:15--12:45&0.8&$+$4.2&1.06&0.14&59&100&Thin cirrus\\
2010 Nov 2&06:45--12:54&1.1&$+$5.2&1.06&0.15&59&102&Photometric\\
2010 Nov 2&07:27--10:32&0.8&$+$5.1&1.06&0.15&59&102&Photometric\\
2010 Nov 3&06:41--13:01&1.1&$+$6.2&1.06&0.15&59&103&Photometric\\
2010 Nov 4&06:39--13:07&1.1&$+$7.2&1.06&0.16&59&104&Thin cirrus\\
2010 Nov 5&07:44--10:45&0.8&$+$8.1&1.06&0.16&59&105&Photometric\\
2010 Nov 6&07:41--10:58&0.8&$+$9.1&1.07&0.16&59&106&Thin cirrus\\
2010 Nov 7&06:48--13:09&1.1&$+$10.2&1.07&0.17&59&107&Photometric\\
2010 Nov 10&07:59--13:11&0.8&$+$13.2&1.07&0.18&58&109&Clouds\\
2010 Nov 12&07:56--12:30&0.8&$+$15.2&1.08&0.19&58&111&Clouds\\
2010 Nov 13&08:09--13:10&0.8&$+$16.2&1.08&0.20&57&112&Clouds\\
2010 Nov 16&08:08--13:08&0.8&$+$19.2&1.09&0.21&56&114&Photometric\\
2010 Nov 26&07:45--12:43&0.8&$+$29.2&1.13&0.26&51&123&Photometric\\
2010 Nov 27&07:48--12:46&0.8&$+$30.2&1.14&0.27&50&124&Photometric\\
2010 Dec 9&07:02--13:19&1.1&$+$42.2&1.21&0.33&42&136&Thin cirrus\\
2010 Dec 10&06:48--08:50&1.1&$+$43.1&1.21&0.34&41&137&Thin cirrus\\
2010 Dec 15&08:57--09:13&1.1&$+$48.1&1.25&0.36&38&143&Clouds\\
2011 Jan 7&04:02--10:51&1.1&$+$71.1&1.42&0.51&25&179&Clouds\\
2011 Jan 8&07:12--07:57&1.1&$+$72.1&1.43&0.52&25&181&Clouds\\
2011 Jan 9&07:09--08:47&1.1&$+$73.1&1.44&0.53&25&183&Clouds\\
2011 Jan 11&04:01--10:50&1.1&$+$75.1&1.45&0.54&24&187&Thin cirrus\\

\enddata
\tablenotetext{a} {All parameters are given for the midpoint of each night's observations, and all images were obtained at Lowell Observatory.}
\tablenotetext{b} {Phase angle.}
\tablenotetext{c} {Position angle of the Sun, measured counterclockwise from north.}
\label{t:obs_summary}
\end{deluxetable}

\begin{deluxetable}{lcccc}  
\tabletypesize{\scriptsize}
\tablecolumns{5}
\tablewidth{0pt} 
\setlength{\tabcolsep}{0.05in}
\tablecaption{Rotation period measurements\tablenotemark{a}.}
\tablehead{   
  \colhead{Average}&
  \colhead{Intervening}&
  \colhead{\# of}&
  \colhead{Period}&
  \colhead{$\sigma$$_{per}$}\\
  \colhead{Date}&
  \colhead{Cycles}&
  \colhead{Pairs}&
  \colhead{(hr)}&
  \colhead{(hr)}
}
\startdata
Aug 13.83&1&1&17.13&--\\
{\it Aug 15.24}&{\it 3}&{\it 8}&{\it 16.67}&{\it 0.17}\\
Aug 15.07&all&9&16.72&0.22\\
\\
Sep 11.63&1&13&17.93&0.78\\
Sep 11.78&3&14&17.20&0.16\\
Sep 11.81&4&8&17.27&0.04\\
{\it Sep 11.79}&{\it 3 or 4}&{\it 22}&{\it 17.23}&{\it 0.13}\\
Sep 11.77&all&35&17.49&0.59\\
\\
Oct 15.39&3&5&18.16&0.20\\
Oct 14.55&5&5&18.13&0.12\\
{\it Oct 15.47}&{\it all}&{\it 10}&{\it 18.15}&{\it 0.15}\\
\\
Nov 03.91&9&6&18.66&0.07\\
\enddata
\tablenotetext{a} {The period we believe is most accurate has been italicized for each observing run.}
\label{t:period_meas}
\end{deluxetable}

\renewcommand{\baselinestretch}{1.0}

\begin{figure}
  \addtocounter{figure}{+1}
  \centering
  \epsscale{1.0}
  \figurenum{1}
  \plotone{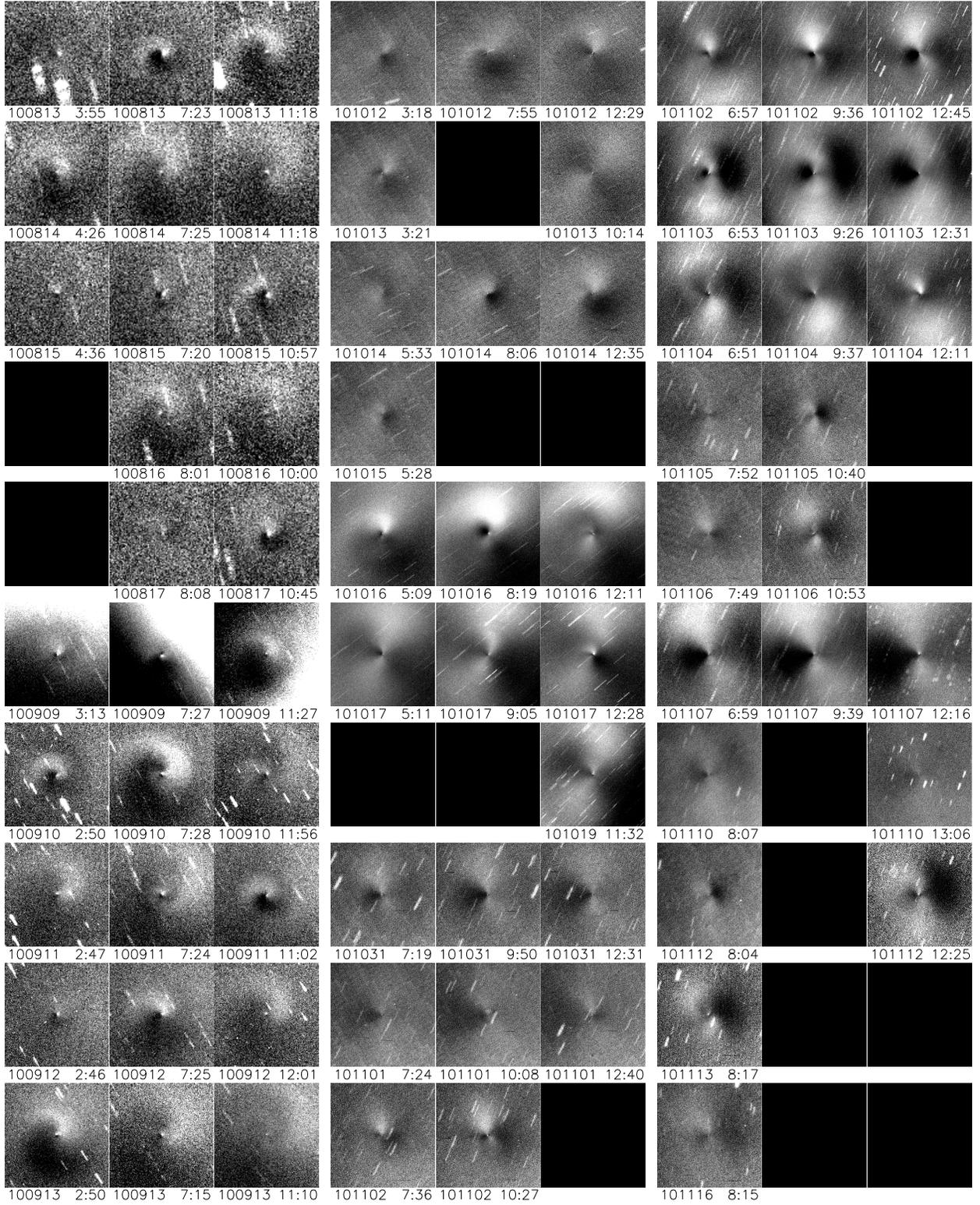}    
  \caption[Representative enhanced CN images]{Continued on next page}
  \label{fig:cn_images}
\end{figure}

\begin{figure}
  \centering
  \epsscale{1.0}
  \figurenum{1}
  \plotone{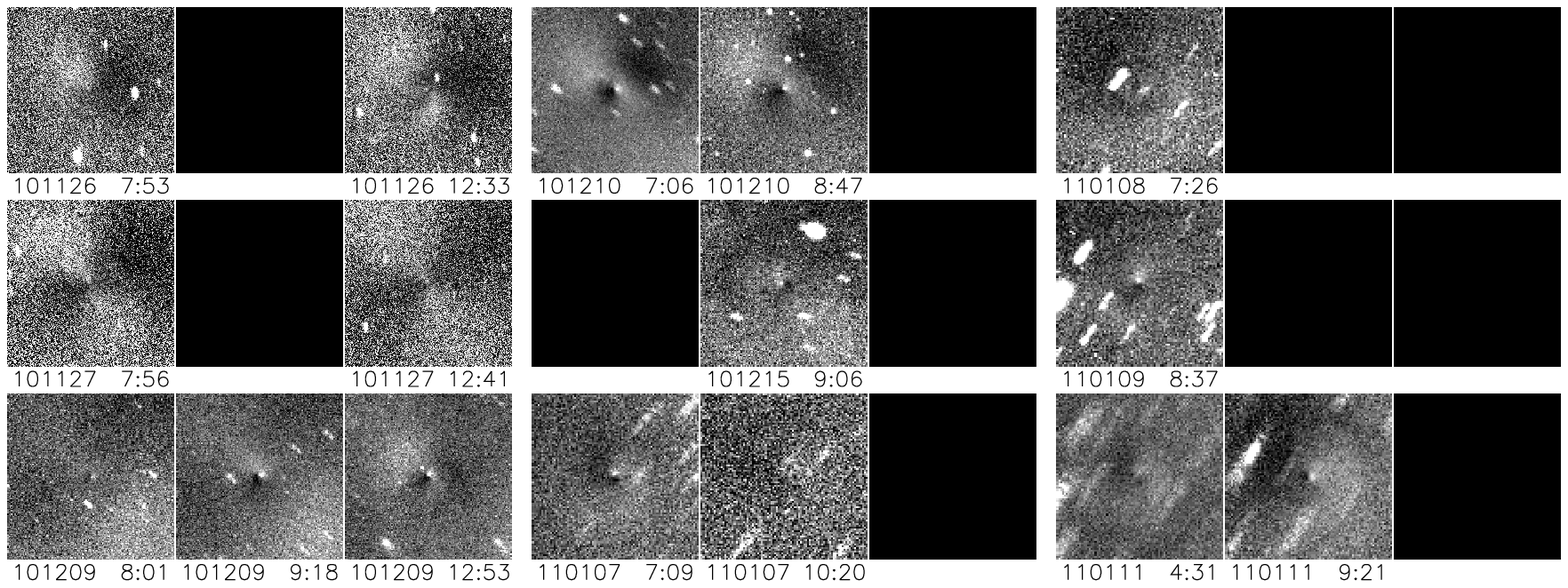}    
  \caption[Representative enhanced CN images]{(Continued from previous page) -- Representative enhanced CN images from each night from 2010 August through 2011 January. Each image was enhanced by dividing by an azimuthal median profile and has the same color stretch, with white representing the brightest areas and black representing the darkest areas. Image dates (YYMMDD) and UT times (HH:MM) are given below each image. Each image has north up and east to the left and is approximately 50,000 km on a side at the comet. Blank images are inserted on nights with less temporal coverage to indicate at which point in a night data were not obtained. Note that the bright white feature which moves from the north to the northwest to the southwest on September 9 was due to a magnitude 3.6 star, not the comet morphology.}
  \label{fig:cn_images}
\end{figure}

\end{document}